\documentclass[aps,prl, twocolumn, showpacs]{revtex4}
\usepackage{graphicx}
\usepackage{dcolumn}
\usepackage{bm}

\begin{document}
\preprint{preprint}
\title{Normal mode splitting and mechanical effects of an optical lattice in a ring cavity}
\author{Julian Klinner}
\email{jklinner@physnet.uni-hamburg.de}
\author{Malik Lindholdt}
\author{Boris Nagorny}
\author{Andreas Hemmerich}
\affiliation{Institut f\"{u}r Laser-Physik, Universit\"{a}t Hamburg, 
Luruper Chaussee 149, 22761 Hamburg, Germany}
\date{\today}

\begin{abstract}
A novel regime of atom-cavity physics is explored, arising when large atom samples dispersively interact with high-finesse optical cavities. A stable far detuned optical lattice of several million rubidium atoms is formed inside an optical ring resonator by coupling equal amounts of laser light to each propagation direction of a longitudinal cavity mode. An adjacent longitudinal mode, detuned by about 3~GHz, is used to perform probe transmission spectroscopy of the system. The atom-cavity coupling for the lattice beams and the probe is dispersive and dissipation results only from the finite photon-storage time. The observation of two well-resolved normal modes demonstrates the regime of strong cooperative coupling. The details of the normal mode spectrum reveal mechanical effects associated with the retroaction of the probe upon the optical lattice.   
\end{abstract}

\pacs{32.80.Pj, 42.50.Vk, 42.50.Pq, 42.65.Sf}

\maketitle

Strongly coupled atom-cavity systems permit studies of the interactions between matter and light on the quantum level with remarkable precision \cite{Ber:94, Mil:05}. Only recently, the significance of mechanical effects in such systems has been recognized and first steps have been made to exploit them for novel cavity-mediated laser cooling (CLC) schemes based upon coherent scattering \cite{Hoo:98, Mun:00, Dom:03, Cha:03, Mau:04}. Previous experiments with high-Q cavities have typically employed standing wave modes, tuned close to an atomic transition, which leads to comparably large spontaneous emission levels. Accessing strong coupling under these circumstances requires a cooperative coupling strength exceeding the cavity linewidth as well as the natural linewidth of the atomic transition. Both conditions together imply the need for very short cavities, which can hold only small atom samples. While this regime is ideal to study interactions between a few atoms and a few photons, a number of exciting scenarios could be envisaged, which would significantly profit from working with large atom samples and very low spontaneous emission levels. One example is the application of CLC, which becomes inherently independent of the internal atomic structure, when operating far from resonance. This bears the perspective to apply CLC to yet unaccessible atomic species, molecules, or atoms at high densities where conventional laser cooling methods cease to work. The combination of a large mode volume with an entirely dispersive atom-cavity coupling should also be well adapted to prepare bound atom-cavity systems involving Bose-Einstein condensates (BECs) (see \cite{Dal:99} for a review), which would be immediately destroyed by excessive spontaneous emission. In such systems the presence of cavity-mediated collective long range forces should give rise to a wealth of novel non-linear quantum phenomena. 
 
These considerations have led us to explore a new regime of the interaction between atoms and high finesse cavities: bound atom-cavity systems involving several million atoms, which operate far from atomic resonances, such that the atom-cavity coupling is entirely of dispersive nature. Only the finite photon storage time remains as a source of dissipation which appears ideal for CLC-schemes with little undesired heating. An indispensable prerequisite is to prove that strong cooperative coupling (SCC) can be achieved in this regime, which implies cavity linewidths in the kHz range, thus imposing particular demands with respect to laser and cavity frequency stability. For atoms in near-resonant standing wave cavities, the SCC-regime can be recognized by the well-known normal mode splitting predicted by the Jaynes Cummings model \cite{Jay:63}. For increasing detuning the atoms decouple from the cavity and the normal mode associated with atomic excitation becomes increasingly difficult to observe, which limits the usefulness of standing wave cavities far from an atomic resonance. In a ring cavity, however, even at large detunings from the atomic resonance a readily observable normal mode splitting arises which results from coherent back-scattering between the two propagation directions of a longitudinal mode. This requires that the atoms are arranged in a Bragg-grating commensurable with the wavelength of the cavity modes, e.g., in a far detuned optical lattice \cite{Gry:01}. 

Recently, we have realized an intra-cavity lattice of several million atoms, using a longitudinal mode of a high finesse ring resonator \cite{Nag:03a}. In ref. \cite{Nag:03b} we could demonstrate the SCC-regime for the case of asymmetric coupling, which manifested itself in optical bistability and self-organization phenomena but did not permit maintaining of a stable lattice. In this Letter, the case of strictly symmetric coupling is explored, which gives rise to a stable optical lattice. In order to show, that this lattice can be operated in the SCC-regime, the normal mode splitting of this bound atom-cavity system is investigated by means of probe transmission spectroscopy. In contrast to previous near-resonant experiments in standing wave cavities (\cite{Rai:89, Zhu:90, Gri:97, Hoo:98, Sau:04, Mau:05}), the interaction of the probe and the lattice beams with the atoms is dispersive, i.e., the atoms practically remain in the internal ground state. When the SCC-regime is accessed, the expected normal mode doublet is observed. As a decisive test of our understanding of the SCC-regime in our system, we analyze the details of the normal mode spectrum, which result from collective mechanical effects associated with the retroaction of the probe upon the optical lattice. Collective mechanical effects have recently received increased attention as a long time disregarded characteristic attribute of atom-cavity systems \cite{Mun:00, Cha:03, Nag:03b}, which, for example, may be utilized to significantly boost the efficiency of CLC-schemes \cite{Zip:04}. The precise control of the normal modes, shown here, represents a significant step forward in the course of implementing efficient CLC, which according to ref. \cite{Els:03} should arise, if the normal mode splitting is tuned to match the vibrational frequency in the optical lattice. 

\begin{figure}
\includegraphics[scale=0.4]{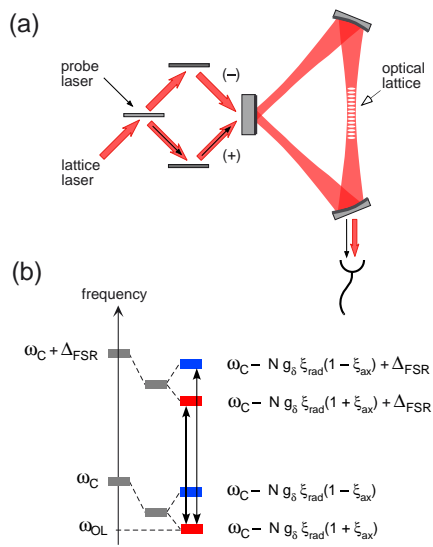}
\caption{\label{Fig1} (a) Sketch of the experimental setup. A far detuned optical lattice (OL) is formed by symmetrically coupling a laser beam (lattice laser) to both propagation directions (denoted by (+) and ($-$)) of some longitudinal cavity mode. A probe beam (thin arrow) is coupled to the (+)-direction of the adjacent longitudinal mode. Inside the cavity the probe and the lattice fields share the same linear polarization perpendicular to the drawing plane. BS = non-polarizing beam splitter, PBS = polarizing beam splitter, HWP = half wave plate. (b) Energy level structure of the atom-cavity system. $\omega_{OL}$ = frequency of the OL, $\omega_c$ = frequency of the corresponding longitudinal mode, if the cavity is empty. The black arrows indicate the probe detunings, for which resonances are expected in the transmission.}
\end{figure} 

The experiment is sketched in Fig.\,\ref{Fig1}(a). An intra-cavity optical standing wave is prepared by symmetrically coupling an identical amount (several microwatts) of light to both propagation directions of a longitudinal cavity mode. The frequency of the light is detuned by $\delta = 0.7$~nm to the red side of the $D_2$-line of $\, ^{85}Rb$ at 780.24~nm. Several $10^6$ Rubidium atoms at a temperature of about $100\,\mu$K are trapped in the corresponding light shift potentials (well depth $\approx 350\,\mu$K), thus forming a stable optical lattice. Loading of the lattice is accomplished by temporarily superimposing a magneto-optic trap upon the cavity mode. The cavity has a round trip path of 97 mm, a finesse of  $F = 1.8\cdot{10^{5}}$ and a field decay rate $\gamma_{c} = \pi \cdot 17.5\,$kHz \cite{Nag:03a}. A high bandwidth stabilization is applied in order to keep one of the lattice beams (($-$)-direction in Fig.\,\ref{Fig1}(a)) in resonance with the cavity \cite{Sch:01}. Heating by spontaneous photons or well depth fluctuations is entirely negligible on the time-scale of several hundred milliseconds. The resonant atom-cavity coupling strength amounts to $g_{0} \equiv \frac{{\omega_{0}}^2}{2 \Gamma}  =  0.67\, \gamma_{c}$ with $\omega_{0}$ and $\Gamma$ being the Rabi-frequency per photon and the atomic spontaneous decay respectively. Hence, for a single atom and resonant light, the cavity operates close to the regime of strong coupling ($g_{0}  > \gamma_{c}$), despite of the relatively large mode radius $w_{0} = 130\,\mu$m. At the large detuning $\delta = 0.7\,$nm, chosen in the experiment, a single atom yields a far smaller coupling $g_{\delta} \equiv  g_{0}/ \sqrt{1 + 4(\frac{\delta}{\Gamma})^2}   \ll \gamma_{c}$, however $N g_{\delta}$, for atom numbers N above a few million, may exceed $\gamma_{c}$, i.e., the lattice in fact can access the SCC-regime. If $|\delta| \gg 0$, $g_{\delta}$ can be approximated as the light shift per photon $\omega_0^2 / 4 \delta$. In order to investigate the normal mode spectrum by means of probe transmission spectroscopy, the adjacent longitudinal mode, detuned by one free spectral range (3.1~GHz) to the high-frequency side of the optical lattice, is used. A diode laser beam (with a few $\mu$W power) is coupled mainly to one propagation direction of the mode ((+)-direction in Fig.\,\ref{Fig1}(a)) and the transmission is recorded versus the frequency. Across the 1 mm axial extension of the lattice the probe phase grating remains commensurable with the optical lattice period. The necessary relative frequency stability of the probe with respect to the resonance frequency of the cavity mode is provided by means of phase locking the probe to the lattice laser. 

The expected normal mode spectrum is readily understood within a simple physical picture (\cite{Els:04}), summarized in Fig.\,\ref{Fig1}(b): if atoms are homogeneously distributed over the mode volume, only forward scattering can arise, such that both degenerate traveling wave modes are shifted in frequency by the same amount $N\, g_{\delta}  \, \xi_{rad}$. In the case of normal dispersion relevant here, this shift is negative. The radial bunching parameter $\xi_{rad} \in [0,1]$, close to unity in this experiment, measures the radial overlap of the atoms with the cavity mode volume. If the atoms are arranged in a lattice with a spatial period of half the optical wavelength $\lambda$, Bragg-scattering in the backwards directions acts to couple both traveling wave modes. This produces an additional splitting, such that the total shift of the two normal modes becomes $N\, g_{\delta}  \, \xi_{rad} (1 \pm \xi_{ax})$, where $\xi_{ax} \in [0,1]$ denotes the axial Debye-Waller factor, which measures the degree of confinement of the atoms within the Bragg grating ($\xi_{ax}$ = 1, if the atoms were point-like particles located with exact $\frac{\lambda}{2}$ spacing). In the SCC-regime the normal modes display a standing wave geometry with the atoms either located in the intensity maxima or minima.  The atom-cavity coupling is either maximized or minimized, which explains the different signs in the corresponding shifts $N\, g_{\delta} \, \xi_{rad} (1 \pm \xi_{ax})$.

\begin{figure}
\includegraphics[scale=0.5]{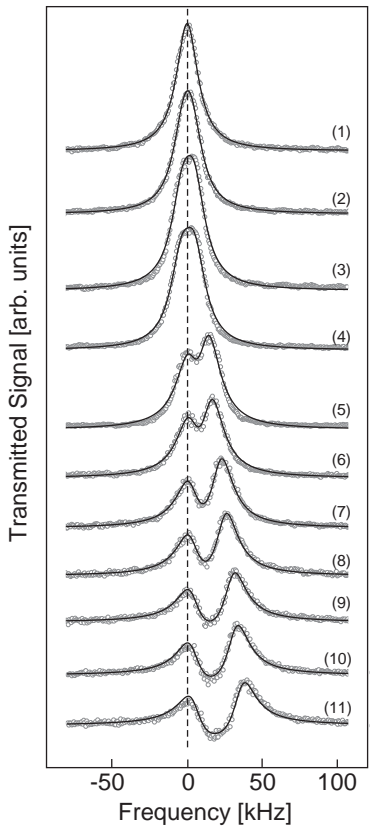}
\caption{\label{Fig2} Normal mode spectra for an optical lattice in a ring cavity. Counting from the upper to the lower trace (1,2,...,11), the number of atoms in the cavity was 0, 0.28, 0.55, 0.83, 1.10,  1.38, 1.66, 1.93, 2.21, 2.48, 2.76 million. Observations are represented by grey circles, while the solid lines are fits by means of the steady state solutions of a semi-classical model.}
\end{figure}

In Fig.\,\ref{Fig2} we show probe transmission spectra for increasing atom numbers in the optical lattice, starting with an empty cavity in the upper trace (trace 1). In the following traces the atom numbers were 0.28, 0.55, 0.83, 1.10,  1.38, 1.66, 1.93, 2.21, 2.48, and 2.76 million respectively. We clearly recognize the predicted doublet structure, when the SCC-regime is entered for particle numbers exceeding about a million. In accordance with the energy budget in Fig.\,\ref{Fig1}(b), the lower frequency component always occurs for the same detuning between the probe laser and the optical lattice (given by $\Delta_{FSR}$), indicated by the dashed line in Fig.\,\ref{Fig2}. The second component is shifted towards higher frequencies with respect to the first by $2 N\, g_{\delta}  \, \xi_{rad} \, \xi_{ax}$. For the largest values of N in Fig.\,\ref{Fig2} (traces (10) and (11)) the signals in between the two resonance peaks fall below the off-resonant level. This can occur because the photodetector recording the transmitted probe also receives the transmission of the copropagating optical lattice beam (cf. Fig.\,\ref{Fig1}), which is decreased, if the optical lattice is perturbed by the probe, as will be discussed below. 

Using eq. 31 in ref. \cite{Gan:00}, the steady-state intra-cavity probe intensities are found to be proportional to  

\begin{eqnarray}
M_{\pm}  \equiv
\left|  \frac{  \sqrt{1 \pm \epsilon}\, (i \delta_{ef} - \gamma_c)  +   i  \sqrt{1 \mp \epsilon  }\,\, g_{ef} \, e^{\pm i \chi} \,     }
{[i(\delta_{ef} - g_{ef} ) - \gamma_c][i(\delta_{ef} + g_{ef} ) - \gamma_c] }\right|^2 \, \, ,  \nonumber \\ & &
\end{eqnarray}

where $\delta_c$ is the detuning between the probe and the empty cavity, $\delta_{ef} \equiv \delta_{c} - N \, g_{\delta} \, \xi_{rad}$ is the effective probe-cavity detuning and $g_{ef} \equiv N \, g_{\delta} \, \xi_{rad} \, \xi_{ax}$ is the effective coupling between the optical lattice and the cavity. The parameter $\epsilon$ denotes the difference between the fractions of the total probe intensity coupled to both propagation directions. In our experiment, because of imperfect polarization optics, only about 96.5 \% of the probe intensity is coupled to the desired direction (the (+)-direction in Fig.\,\ref{Fig1}(a)) while the rest (3.5 \%) is fed to the counterpropagating  ($-$)-mode, i.e., $\epsilon$ is about 0.93. As a consequence, a fraction of the probe power circulating in the cavity gives rise to a shallow optical standing wave even if $g_{ef} = 0$. The parameter $\chi$ denotes the relative spatial phase between the atomic grating and this probe light grating at $g_{ef} = 0$. If $\epsilon  = 1$, i.e., for ideally single-sided coupling, $\chi$ drops out of eq.(1) and a fully symmetric spectral resonance doublet is predicted. For $\epsilon < 1$, the two spectral components acquire different weights depending on the value of $\chi$. The reason is that the normal modes are excited via both propagation directions with different phases depending on the value of $\chi$, which can yield constructive interference for one and destructive interference for the other normal mode. 

For a detailed understanding of the normal mode spectra of Fig.\,\ref{Fig2}, the retroaction of the probe grating upon the optical lattice has to be accounted for. For large values of $g_{ef}$, the probe is strongly reflected by the atom grating. This gives rise to a significant force upon the atoms, which needs to be balanced by the lattice beams. Momentum conservation in the ring cavity, a direct implication of the translational invariance of the system, thus requires a power imbalance to be established between the intra-cavity lattice beams, in favor of the one counterpropagating with respect to the probe. In the experiment this lattice beam is kept at maximum incoupling by a fast servo lock, such that the lattice can only react by a power decrease of the copropagating lattice beam. This self-organized power adjustment is accomplished via a spatial shift of the lattice, which tunes the copropagating lattice mode slightly out of resonance with the cavity, such that less power is in-coupled. Note that interference between the probe and the lattice fields does not play a role for the atom dynamics, because of their 3.1~GHz frequency difference. Electronic retroaction of the probe upon the lattice beams via the servo-lock is carefully avoided. Hence, it appears to be an obvious assumption that the power reduction of the copropagating lattice beam is proportional to the reflected probe power,  which according to eq.(1) is proportional to $M_{-}$. With this in mind we model the transmission spectra of Fig.\,\ref{Fig2} by $M(\delta_{ef}, g_{ef}, \chi, R, S) \equiv S \cdot  [M_{+}(\delta_{ef}, g_{ef}, \chi,\,\epsilon=0.93) - R \cdot M_{-}(\delta_{ef}, g_{ef}, \chi, \epsilon=0.93)]$ with $M_{\pm}$ being the expressions of eq.(1), and $g_{ef}$, $\chi$, $R$ and $S$ taken as fit parameters. 

\begin{figure}
\includegraphics[scale=0.35]{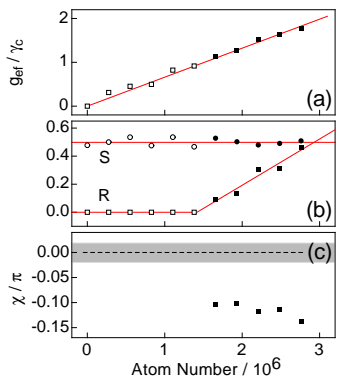}
\caption{\label{Fig3} Plot of the effective coupling strength $g_{ef}$ (a), the retroaction parameter $R$ (b, lower trace), the scaling factor $S$ (b, upper trace), and the spatial phase of the atom grating $\chi$ (c) versus the particle number in the optical lattice. The grey area in (c) indicates the range of possible values of $\chi$ for $g_{ef} = 0$. }
\end{figure}

The fits derived from our model are plotted as solid lines in Fig.\,\ref{Fig2}, the corresponding values of the fit parameters $g_{ef}$, $\chi$, $R$ and $S$ are shown in Fig.\,\ref{Fig3}. The values found for $g_{ef}$, plotted in (a), display the expected linear dependence upon the particle number N, which serves as a consistency check for the fit procedure. In (b) the retroaction parameter $R$ (lower trace), which accounts for the perturbation of the optical lattice by the probe, and the scaling factor $S$ (upper trace) are plotted. The scaling factors $S$ coincide for all fits to better than 12 \%, which shows that the relative signal sizes are correctly predicted by our model. For the first data points in (a) and (b), corresponding to N=0, $g_{ef}$ is trivially zero, and $\chi$, and $R$ drop out of the fit model $M(\delta_{ef}, g_{ef}, \chi, R, S)$, which becomes a simple Lorentzian with $17.5\,kHz$ linewidth. For traces (2)-(6), which yield the open symbols in Fig.\,\ref{Fig3}, we find $R=0$. For these traces, characterized by $g_{ef} < \gamma_c$, the dependence of $M(\delta_{ef}, g_{ef}, \chi, R, S)$ upon $\chi$ turns out to be very weak, such that no reliable value of $\chi$ can be specified. This is not surprising, since for low $g_{ef}$ the spectra display only little structure. The optical paths between the beam splitter, directing the laser beams to the ($\pm$)-modes of the cavity, and the intra-cavity atomic sample differ by less than $\pm 2\%$ (cf. Fig.\,\ref{Fig1}(a)).  Hence, for $g_{ef} = 0$ we may expect $\chi$ to lay within the grey area in Fig.\,\ref{Fig3}(c). The filled symbols in Fig.\,\ref{Fig3} result from traces  (7)-(11), which satisfy $g_{ef} > \gamma_c$. In this regime $R$ grows linearly with N, and we find the values of $\chi$ shown in (c). The deviation of $\chi$ from zero, which corresponds to a shift of the atom grating, nicely confirms the presence of retroaction. A more complete model of the retroaction of the probe upon the optical lattice has to be left to a forthcoming publication. Nevertheless, the excellent agreement between the data and the fits in Fig.\,\ref{Fig2} indicates, that our simple model correctly reproduces the basic essence of this effect.

In summary, we have explored a new regime of atom-cavity physics, arising when large atom samples dispersively interact with high-finesse optical cavities. A stable far-detuned optical lattice is formed in a high-Q ring cavity relying on a dispersive atom-cavity coupling. The regime of strong cooperative coupling is demonstrated by spectroscopically resolving a normal mode doublet. The details of the normal mode spectrum reveal mechanical effects associated with the retroaction of the probe upon the optical lattice. Our system should permit the implementation of cavity-mediated sideband cooling. The regime of dispersive atom-cavity coupling may prove appropriate for bringing together the worlds of cavity quantum electrodynamics and quantum degenerate gases.

\begin{acknowledgments}
This work has been supported by the Deutsche Forschungsgemeinschaft under contract He2334/6-1. 
\end{acknowledgments}

\end{document}